\documentclass[a4paper,10pt,notitlepage]{article}

\usepackage[latin1]{inputenc}
\usepackage{amssymb, bm}
\usepackage{graphicx}
\usepackage{lscape}
\usepackage{setspace}
\doublespace
\usepackage{lineno}
\linenumbers

\def\0{\phantom{0}}
\def\.{\phantom{.}}

\title{Comment on ``An optimized potential for carbon dioxide'' [J. Chem. Phys. 122, 214507 (2005)]}
\author{Thorsten Merker, Jadran Vrabec\footnote{ Author to whom correspondence should be adressed: Tel.: +49-711/685-66107, Fax:
+49-711/685-66140, Email: vrabec@itt.uni-stuttgart.de}, and Hans Hasse}
\date{Institut f\"ur Technische Thermodynamik und Thermische Verfahrenstechnik, Universit\"at Stuttgart, 70550 Stuttgart, Germany}

\begin{document}
\maketitle
In a recent article, Zhang and Duan \cite{zhang2005} presented a new potential model for carbon dioxide ($\mathrm{CO_2}$). It consists of three Lennard-Jones (LJ) sites to account for repulsion and dispersion and three distributed partial charges to describe the quadrupolar interaction. The molecular model is rigid and rotationally symmetric around the molecular axis. In that work \cite{zhang2005}, simulation results on vapor-liquid equilibria (VLE), radial distribution function and self-diffusion coefficient have been reported for the new $\mathrm{CO_2}$ model which are in excellent agreement with experimental data. They also compared to results from different other models from the literature. The new model is found "to be superior to the previous models in general".

For the VLE properties, deviations between model and experimental data are reported to be \cite{zhang2005}:  0.7~\% for vapor pressure,  0.1~\% for saturated liquid density, 2.3~\% for saturated vapor density, and 1.9~\% for heat of vaporization. 

Particularly for vapor pressure, which is the most sensitive of those properties, the authors claim to achieve an average accuracy of better than 1~\% over the entire temperature range with a molecular model which is noteworthy for any molecular model.

As there is widespread scientific interest in $\mathrm{CO_2}$, the model by Zhang and Duan \cite{zhang2005} was employed for subsequent work \cite{zhang2007,Eslami2007,idrissi2007,idrissi2006,zhang2006}. It should be pointed out that no VLE data were published in \cite{zhang2007,Eslami2007,idrissi2007,idrissi2006,zhang2006}. We recently tested the $\mathrm{CO_2}$ model from Zhang and Duan \cite{zhang2005} and found results that strongly deviate from these reported by Zhang and Duan \cite{zhang2005}. The new simulation results also strongly deviate from experimental data, particularly the vapor pressure and the saturated vapor density.

The present assessment was made on the basis of two simulation tools that employ different methods to determine VLE. Firstly, the Grand Equilibrium method \cite{jadran} was used as implemented in our simulation tool $ms2$ \cite{deublein} and, secondly, the Gibbs ensemble \cite{Gibbs} was used as implemented in the freely available simulation tool TOWHEE \cite{TOWHEE}. Both programs have proven to be correct e.g. in the recent Industrial Fluid Property Simulation Challenge for the case of ethylene oxide which is very similar to the present case \cite{eckl}.

For our approach with $ms2$, molecular dynamics (MD) simulations were performed in the liquid phase containing 1024 molecules. After a sufficient equilibration period, the chemical potential was calculated by Widoms's insertion method \cite{widom} over 300 000 time steps. According to the Grand Equilibrium method \cite{jadran}, the dew point was sampled with Monte-Carlo (MC) simulations where approximately 500 molecules were used. The simulation details were similar to those published in \cite{eckl} and are not repeated here.

With TOWHEE \cite{TOWHEE}, Gibbs ensemble MC simulations \cite{Gibbs} were performed. There, smaller systems were studied, containing 500 molecules. After an equilibration over 5 000 loops without volume and molecules transfer moves, followed by 10 000 loops with these moves, 50 000 production loops were performed. Other simulation details were similar to those published in \cite{ozgur} and are not repeated here.

The Present simulation results are compared to those reported by Zhang and Duan \cite{zhang2005} in Table 1 as well as in Figures 1 and 2, which also contains results from a highly accurate reference data for $\mathrm{CO_2}$ \cite{data} recommended by the National Institute of Science and Technology \cite{NIST}. Figure 2 shows deviation plots where it can be seen that the present data sets based on the two different simulation methods agree with each other within their (combined) error bars throughout. However, they are significantly off the data by Zhang and Duan \cite{zhang2005} which coincide excellently with the experiment.

The present simulation data (Grand Equilibrium) show the following average deviations from experimental data: 18~\% for vapor pressure, 0.6~\% for saturated liquid density, 17~\% for saturated vapor density and 4.6~\% for heat of vaporization. Only the saturated liquid density by Zhang and Duan \cite{zhang2005} and the heat of vaporization are in good agreement with the experimental data, the vapor pressure and saturated vapor density are significantly too high throughout most of the temperature range.

It has to be concluded that the $\mathrm{CO_2}$ model by Zhang and Duan \cite{zhang2005} is not generally superior to previous models.

\clearpage



\newpage
\begin{landscape}
\begin{table}[ht]
\noindent
\caption{Vapor-liquid equilibria of carbon dioxide: present simulation results with the Grand Equilibrium method (GE) \cite{jadran} and the Gibbs ensemble (Gibbs) \cite{Gibbs} compared to simulation results by Zhang and Duan (Zhang) \cite{zhang2005} and the reference EOS (eos) \cite{data}. The number in parentheses indicates the statistical uncertainty in the last digit.}
\label{tab_vle1}
\footnotesize{
\medskip
\begin{center}
\begin{tabular}{c|cccc|cccc|cccc|cccc} \hline\hline
$T$ & $p_{\mathrm{GE}}$ & $p_{\mathrm{Gibbs}}$ & $p_{\mathrm{Zhang}}$ & $p_{\mathrm{eos}}$ & $\rho'_{\mathrm{GE}}$ & $\rho'_{\mathrm{Gibbs}}$ & $\rho'_{\mathrm{Zhang}}$ & $\rho'_{\mathrm{eos}}$ & $\rho''_{\mathrm{GE}}$ & $\rho''_{\mathrm{Gibbs}}$ & $\rho''_{\mathrm{Zhang}}$ & $\rho''_{\mathrm{eos}}$ & $\Delta h^{\mathrm{v}}_{\mathrm{GE}}$ & $\Delta h^{\mathrm{v}}_{\mathrm{Gibbs}}$ & $\Delta h^{\mathrm{v}}_{\mathrm{Zhang}}$ & $\Delta h^{\mathrm{v}}_{\mathrm{eos}}$  \\
K & MPa & MPa & MPa & MPA &  mol/l & mol/l & mol/l & mol/l &  mol/l & mol/l & mol/l & mol/l &  kJ/mol & kJ/mol & kJ/mol & kJ/mol\\ \hline
220 & 0.77(1)  & ---    & ---        & 0.599 & 26.42(3)  & ---    & ---         & 26.497 & 0.466(1) & ---    & ---      & 0.359 & 14.41\0(1)  & ---    & ---         & 15.179\\
230 & 1.08(1)    & ---    & 0.90\0(2)  & 0.893 & 25.48(2)  & ---    & 25.59\0(1)  & 25.646 & 0.641(2)   & ---    & 0.558(7) & 0.529 & 13.717(6)       & ---    & 13.90(5)    & 14.437\\
240 & 1.50(2)  & ---    & 1.29\0(1)  & 1.283 & 24.60(2)  & ---    & 24.697(6)   & 24.742 & 0.882(2) & ---    & 0.79\0(2)& 0.757 & 12.994(7)  & ---    & 13.19(6)    & 13.628\\
250 & 2.11(2)  & ---    & 1.788(2)   & 1.785  & 23.66(3)  & ---    & 23.740(8)   & 23.767 & 1.253(4) & ---    & 1.09\0(3)& 1.06\0 & 12.138(8)  & ---    & 12.40(6)    & 12.733\\
260 & 2.86(2)  & 2.9(2) & 2.41\0(2)  & 2.419 & 22.58(4)  & 22.6(2)& 22.68\0(1)  & 22.697 & 1.738(6)& 1.8(1) & 1.48\0(4)& 1.464 & 11.14\0(1)  & 11.2(1)& 11.51(5)    & 11.728\\
270 & 3.69(2)  &3.8(3)    & 3.18\0(3)  & 3.203 & 21.33(6)  & 21.3(3)    & 21.50\0(1)  & 21.491 & 2.29\0(1) & 2.4(3)    & 2.01\0(5)& 2.008 & 10.06\0(1) & 10.0(2)    & 10.46(4)    & 10.569\\
280 & 4.78(3)  & 4.8(2) & 4.11\0(5)  & 4.161 & 19.9\0(1) & 20.1(2)& 20.09\0(2)  & 20.077 & 3.14\0(2)& 3.2(2) & 2.76\0(4)& 2.766 &\08.71\0(2) &\08.8(1)& \09.17(3)   & \09.183\\
290 & 5.92(4)  & ---    & 5.23\0(6)  & 5.318  & 18.1\0(4)   & ---    & 18.29\0(4)  & 18.284 & 4.14\0(4)      & ---    & 3.96\0(6)& 3.907 &\07.16\0(4)       & ---    & \07.45(4)   & \07.399\\
\hline\hline
\end{tabular}
\end{center}
}
\end{table}
\end{landscape}
\clearpage

\newpage
\listoffigures


\newpage
\begin{figure}[ht]
\includegraphics[scale=0.7]{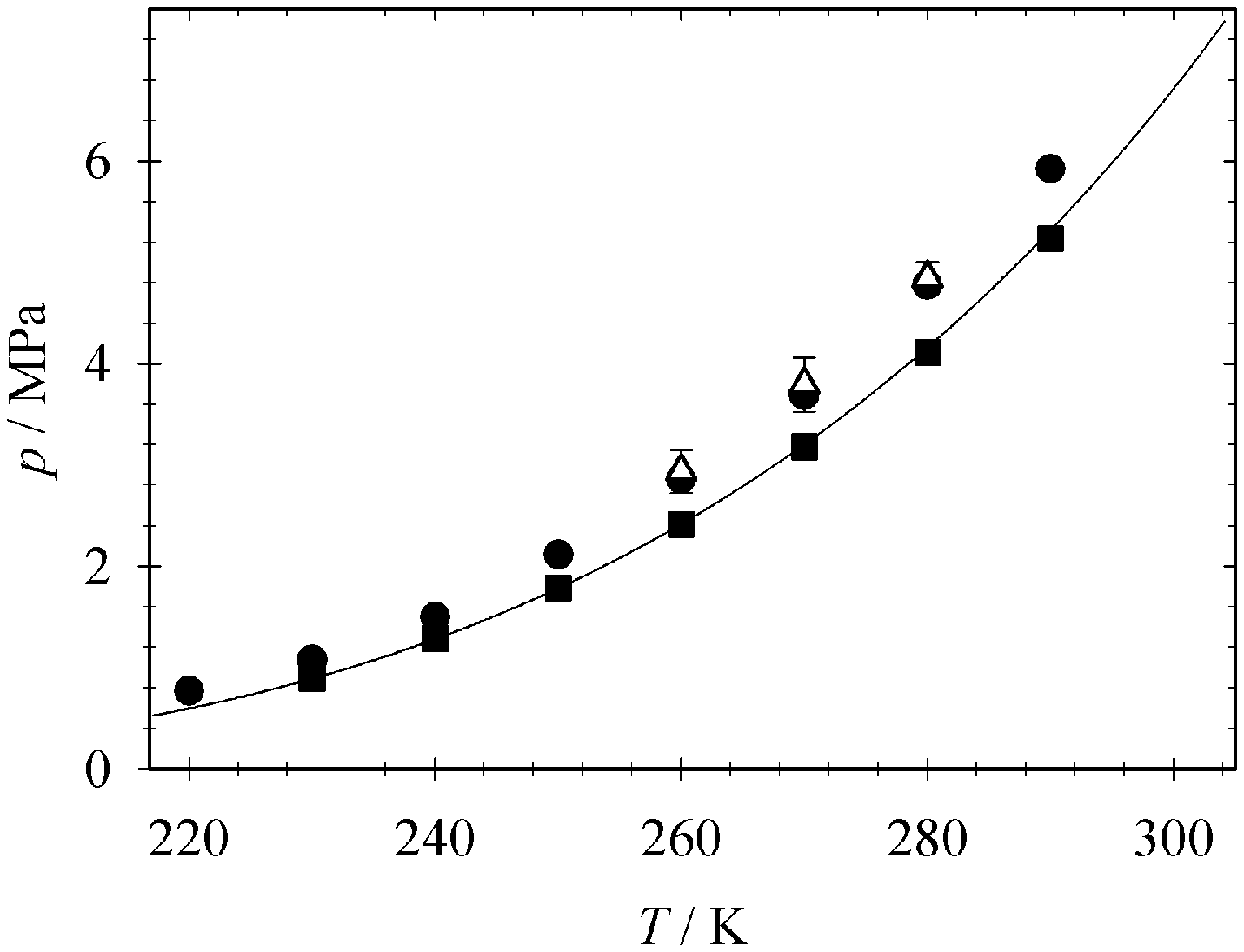}
\bigskip
\caption[Vapor pressure of carbon dioxide. {\large $\circ$}~present results Grand Equilibrium method \cite{jadran}, {\small $\triangle$}~present results Gibbs ensemble method \cite{Gibbs}, {\small $\square$}~results by Zhang and Duan \cite{zhang2005}, {---}~reference EOS \cite{data}.]{Merker et al.\label{fig_vle_p}}
\end{figure}


\newpage
\begin{figure}[ht]
\includegraphics[scale=0.7]{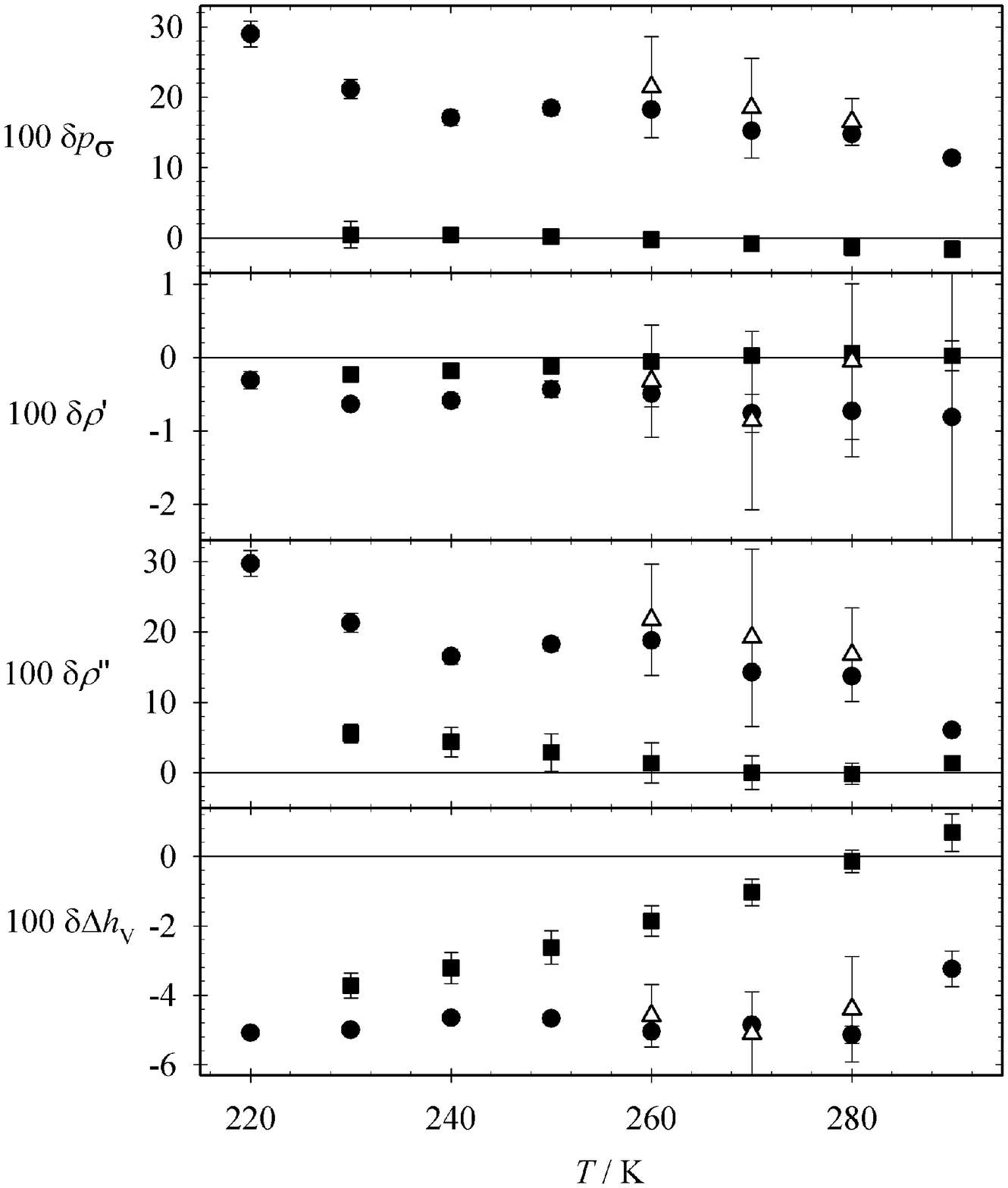}
\bigskip
\caption[Relative deviations of vapor-liquid equilibrium properties between simulation and reference EOS \cite{data} ($\delta z = (z_{\mathrm{sim}} - z_{\mathrm{eos}}) / z_{\mathrm{eos}}$). {\large $\circ$}~present results Grand Equilibrium method \cite{jadran}, {\small $\triangle$}~present results Gibbs ensemble method \cite{Gibbs}, {\small $\square$}~results by Zhang and Duan \cite{zhang2005}, {---}~reference EOS \cite{data}.]{Merker et al.\label{fig_vle_abw}}
\end{figure}


\begin{thebibliography}{00}

\bibitem{zhang2005}
Z. Zhang, Z. Duan,
\newblock An optimized molecular potential for carbon dioxide
\newblock \emph{J. Chem. Phys}, 122 (2005) 214507.

\bibitem{zhang2007}
ZG. Zhang, ZH. Duan,
\newblock Moleular level simulations of the H2O, CO2 and CO2-H2O systems up to high temperatures and pressures
\newblock \emph{Geochemica et Cosmochima Acta}, 71 (2007) A1164.

\bibitem{Eslami2007}
H. Eslami, F. Müller-Plathe,
\newblock Molecular dynamics simulation of sorption of gases in polystyrene
\newblock \emph{Macromolecules}, 40 (2005) 6413-6421.

\bibitem{idrissi2007}
A. Idrissi, P. Damay, M. Kiselev,
\newblock Nearest neighbor assessments of spatial distribution of CO2: A molecular dynamics analysis
\newblock \emph{Chem. Phys.}, 332 (2007) 139-143.

\bibitem{idrissi2006}
A. Idrissi, S. Longelin, P. Damay,
\newblock Analysis of the transverse and the longitudinal pseudodiffusion of CO2 in sub- and supercritical states: A molecular-dynamics analysis
\newblock \emph{J. Chem. Phys.}, 125 (2006) 224501.

\bibitem{zhang2006}
ZG. Zhang, ZH. Duan,
\newblock Equation of state of the H2O, CO2 and H2O-CO2 systems up to 10 GPa and 2573.15 K: Molecular dynamics simulation with ab initio potential surface
\newblock \emph{Geochemica et Cosmochima Acta}, 70 (2006) 2311-2324.

\bibitem{jadran}
J. Vrabec, H. Hasse,
\newblock Grand Equilibrium: Vapor-liquid Equilibira by a New Molecular Simulation Method
\newblock \emph{Mol. Phys.}, 100 (2002) 3375-3383.

\bibitem{deublein}
S. Deublein, B. Eckl, J. Stoll, S. V. Lishchuk, M. Bernreuther, J. Vrabec, H. Hasse,
\newblock $ms2$: A Molecular Simulation Code for Thermodynamic Properties
\newblock In preparation, (2008).

\bibitem{Gibbs}
A. Z. Panagiotopoulos,
\newblock Dicrect determination of phase coexistence properties of fluids by Monte Carlo simulation in a new ensemble
\newblock \emph{Mol. Phys.}, 61 (1987) 813-826.

\bibitem{TOWHEE}
Towhee Monte Carlo molecular simulation code,
\newblock http://towhee.sourceforge.net.

\bibitem{eckl}
B. Eckl, J. Vrabec, H. Hasse,
\newblock On the Application of Force Fields for Predicting a Wide Variety of Properties: Ethylene Oxide as an Example
\newblock \emph{Fluid Phase Equilib.}, (2008), doi:10.1016/j.fluid.2008.02.002.


\bibitem{widom}
B. Widom,
\newblock Some Topics in the Theory of Fluids
\newblock \emph{J. Chem. Phys.}, 39 (1963) 2808-2812.

\bibitem{ozgur}
A. Özgür Yazaydin, M. G. Martin,
\newblock Bubble point pressure estimates from Gibbs ensemble simulations
\newblock \emph{Fluid Phase Equilib.}, 260 (2007) 195-198.


\bibitem{data}
R. Span, W. Wagner,
\newblock A New Equation of State for Carbon Dioxide Covering the Fluid Region from Triple-Point Temperature to 1100K at Pressures up to 800 MPa.
\newblock \emph{J. Phys. Chem. Ref. Data}, 25 (1996) 1509-1596.

\bibitem{NIST}
National Institute of Standards and Technology,
\newblock http://www.nist.gov.



\end{thebibliography}
\end{document}